\documentclass[11pt]{article}
\usepackage{makeidx}
\usepackage{fleqn}
\usepackage{epsfig}
\input{amssym.def}
\input{amssym.tex}

\author{ }
\date{ }
\makeatletter
\def\Egg{${\rm E}_{\gamma\gamma}$}
\def\MH{${\rm M}_{\mbox{\tiny{H}}}$}

\def\MA{${\rm M}_{\mbox{\tiny{A}}}$}

\def\lsim{\compoundrel<\over\sim}
\def\compoundrel#1\over#2{\mathpalette\compoundreL{{#1}\over{#2}}}
\def\compoundreL#1#2{\compoundREL#1#2}
\def\compoundREL#1#2\over#3{\mathrel
      {\vcenter{\hbox{$\m@th\buildrel{#1#2}\over{#1#3}$}}}}
\begin{document}
\begin{flushright}
hep-ph/0403303\\
Freiburg--THEP 04/04\\
KEK Preprint 2004-5
\end{flushright}
\vspace*{3.5cm}
\begin{center}
{\Huge Higgs~~Self-Coupling~~in~~$\gamma\gamma$~~Collisions}\\
\vspace*{5cm}
{\Large R. BELUSEVIC\,$^{a}$~~and~~G. JIKIA\,$^{b}$}\\
\vspace*{1.5cm}
$^{a}\,${\large\em KEK, Oho 1-1, Tsukuba-shi, Ibaraki-ken 305-0801, Japan}\\ 
\vspace*{0.3cm}
$^{b}\,${\large\em Albert--Ludwigs--Universit\"{a}t Freiburg,
            Fakult\"{a}t f\"{u}r Mathematik und Physik\\
       Hermann--Herder Str. 3a, D-79104 Freiburg, Germany}
\end{center}

\newpage

\vspace*{8cm}
\begin{center}
\begin{minipage}[t]{10.6cm}
``{\large\em The changing of bodies into light, and light into bodies, is very
conformable to the course of nature, which seems delighted with
transmutations.}"

\addvspace{3mm}
\hfill\ {\large Isaac Newton}
\end{minipage}
\end{center}

\newpage

\begin{center}
{\Large\bf HIGGS~~SELF-COUPLING~~IN~~$\gamma\gamma$~~COLLISIONS} 
\end{center}

\vspace*{1cm}
\begin{center}
\begin{minipage}[t]{13cm}
{\small\bf Abstract\,:}\hspace*{2mm}{\small To establish the Higgs mechanism
experimentally, one has to determine the Higgs self-interaction potential 
responsible for the electroweak symmetry breaking. This requires a measurement
of the trilinear and quadrilinear self-couplings of the Higgs particle, as
predicted by the Standard Model (SM). We propose to measure the trilinear Higgs
self-coupling in $\gamma\gamma$ collisions just above the kinematic threshold 
${\rm E}_{\mbox{\scriptsize{thr}}} = 2{\rm M}_{\mbox{\tiny{H}}}$, where 
M$_{\mbox{\tiny{H}}}$ is the Higgs mass. Our calculation reveals that the
sensitivity of the cross-section $\mbox{\large{$\sigma$}}_{\gamma\gamma\,
\rightarrow\,\mbox{\tiny{HH}}}$ to the Higgs self-coupling is maximal near the 
$2{\rm M}_{\mbox{\tiny{H}}}$ threshold for ${\rm M}_{\mbox{\tiny{H}}} =
115-150$ GeV, and is {\em larger} than the sensitivities of $\mbox{\large 
{$\sigma$}}_{e^{+}e^{-}\,\rightarrow\,\mbox{\tiny{$Z$HH}}}$ and $\mbox{\large
{$\sigma$}}_{e^{+}e^{-}\,\rightarrow\,\nu\bar{\nu}\mbox{\tiny{HH}}}$ to this
coupling for $2{\rm E}_{e} \leq 700$ GeV. We envisage to (a) study $\gamma\,+\,
\gamma\,\rightarrow\,{\rm H}$ by constructing an X-band $e^{-}e^{-}$ linac and
a terawatt laser system in order to produce Compton-scattered $\gamma$-ray
beams for a 160-GeV photon collider ($2{\rm E}_{e} = 200$ GeV); (b) add a
positron source and repeat all measurements done at LEP and SLC with much
better precision; and (c) subsequently install 70-MeV/m rf cavities in order to
study $e^{+}\,+\,e^{-}\,\rightarrow\,{\rm H}\,+\,Z$, $e^{+}\,+\,e^{-}\,
\rightarrow\,t\bar{t}$ and $\gamma\,+\,\gamma\,\rightarrow\,{\rm H}\,+\,
{\rm H}$ at $2{\rm E}_{e} \lsim 350$ GeV. The total length of the linac would
be about 7 km.} 

\end{minipage}
\end{center}

\vspace*{0.5cm}
\renewcommand{\thesection}{\arabic{section}}
\section{Introduction}
\vspace*{0.8cm}

\setcounter{equation}{0}

~~~~Enormous progress has been made in the field of high-energy, or elementary
particle, physics over the past four decades. The existence of a subnuclear
world of quarks and leptons, whose dynamics can be described by quantum field
theories possessing gauge symmetry ({\small\bf gauge theories}) has been firmly
established. The {\small\bf Standard Model} (SM) of particle physics gives a 
coherent quantum-mechanical description of electromagnetic, weak and strong
interactions based on fundamental constituents --- quarks and leptons ---
interacting via force carriers --- photons, $W$ and $Z$ bosons, and gluons. 

In this model, the relativistic theory of photons and electrons (called quantum
electrodynamics, or QED) is a consequence of a {\small\bf spontaneously broken
symmetry} in a theory in which the weak and electromagnetic interaction are
initially unified and the corresponding force carriers ({\small\bf gauge 
bosons}) are massless. To account for the observed mass spectrum of the field
quanta, one postulates the existence of a {\small\bf Higgs field}, which is a
scalar under spatial rotations but is a weak isodoublet. Like the graviton and
the gauge bosons of the Standard Model, the Higgs boson mediates a fundamental
force of nature. The coupling of the Higgs field to the vector fields that 
mediate the electroweak interaction is arranged so as to give the $W$ and $Z$
masses in the $10^{2}$ GeV range, while maintaining the photon mass at zero.
The Higgs field thus provides the mechanism for electroweak symmetry breaking.

All of the couplings of the Higgs boson to fermions and gauge bosons are
completely determined in terms of coupling constants and fermion masses. The
coupling of a fermion to the scalar Higgs field is proportional to the mass of
the fermion. The Higgs boson mass represents a free parameter of the model.

The $W$- and $Z$-boson masses are related by the electroweak mixing angle 
$\theta_{\mbox{\tiny{W}}}$ (also called the Weinberg angle): ${\rm M}_{\mbox
{\tiny{W}}} = {\rm M}_{\mbox{\tiny{Z}}}\cos\theta_{\mbox{\tiny{W}}}$. The fact
that interactions of all gauge bosons are determined by the electric charge and
one free parameter, $\theta_{\mbox{\tiny{W}}}$, means that the Standard Model
is a (partially) unified theory of the weak and electromagnetic interactions. 

The ${\rm SU}(2)_{\mbox{\tiny{L}}\!}\otimes\!{\rm U}(1)_{\mbox{\tiny{Y}}}$ 
gauge invariance of the SM requires masses of the gauge bosons to be zero,
since the presence of a mass term ${\rm M}^{2}A_{\mu}A^{\mu}$ for these 
particles would render the model non-invariant under the gauge transformation
$A_{\mu} \rightarrow A_{\mu} - \partial_{\mu\,}\chi (x)$. In order to provide a
mechanism for the generation of particle masses without violating the gauge 
invariance of the model, a complex scalar SU(2) doublet with four real fields
and {\em hypercharge} ${\rm Y} = 1$ is introduced:
\begin{equation}
\Phi \,\equiv \left (\!\!
\begin{array}{c}
\phi^{+} \\*[0.7mm]
\phi^{0}
\end{array} \!\!\right )
\end{equation}
in analogy with the $K^{0}$ system. The reason we need two complex fields
rather than one is that three degrees of freedom are required to generate the
masses of three gauge fields (the $W^{\pm}$ and the $Z^{0}$). The remaining
degree of freedom will show up as a neutral massive Higgs field. The Higgs
doublet conveniently serves two purposes, for it can also give mass to the
fermions. The dynamics of the field $\Phi$ is described by the Lagrangian
\begin{equation}
{\cal L}_{\Phi} \,=\, (D_{\mu}\Phi )^{\dag}(D^{\mu}\Phi ) \,-\, V(\Phi ) 
\end{equation}
where $(D_{\mu}\Phi )^{\dag}(D^{\mu}\Phi )$ is the kinetic-energy term of
${\cal L}_{\Phi}$ and 
\begin{equation}
V(\Phi^{\dag}\Phi ) \,=\, \mu^{2\,}\Phi^{\dag}\Phi \,+\, \lambda\mbox{\large
{$($}}\Phi^{\dag}\Phi\mbox{\large{$)$}}^{2}
\label{eq:potential}
\end{equation}
is the Higgs self-interaction potential with a minimum at 
\begin{equation}
\langle\Phi\rangle_{0}^{~} \,=\, \left (\!\!
\begin{array}{c}
0 \\*[0.7mm]
v/\sqrt{2}
\end{array} \!\!\right ),~~~~~~~~~v \,=\, \sqrt{-\mu^{2}/\lambda}
\end{equation}
The quantity v is the {\small\bf vacuum expectation value} of $\Phi$. In
the {\small\bf unitary gauge},
\begin{equation}
\Phi \,=\, \frac{\raisebox{-.3ex}{\mbox{\small{1}}}}{\sqrt{\mbox{\small
{2}}}}\left (\!\!
\begin{array}{c}
0 \\*[0.7mm]
v + {\rm H}
\end{array} \!\right ) 
\end{equation}
where H is the physical {\small\bf Higgs field}. The potential
$V(\Phi )$ gives rise to terms involving only the scalar field H:
\begin{equation}
V_{\mbox{\tiny{H}}} \,=\, \frac{\raisebox{-.3ex}{\mbox{\small{1}}}}
{\raisebox{.2ex}{\mbox{\small{2}}}}\mbox{\Large{$($}}\mbox{\small{2}}\lambda
v^{2}\mbox{\Large{$)$}}{\rm H}^{2} \,+\, \lambda v\,{\rm H}^{3} \,+\, \frac
{\raisebox{-.3ex}{$\lambda$}}{\raisebox{.2ex}{\mbox{\small{4}}}}\,{\rm H}^{4}
\end{equation} 
From this we infer that the {\small\bf Higgs mass} 
\begin{equation}
{\rm M}_{\mbox{\tiny{H}}} \,=\, \sqrt{\mbox{\small{2}}\lambda}\,v 
\end{equation}
is related to the quadrilinear self-coupling strength $\lambda$. The {\small\bf
trilinear self-coupling} of the Higgs field is given by 
\begin{equation}
\lambda_{\mbox{\tiny{HHH}}} \,\equiv\, \lambda v \,=\, \frac{{\rm M}_{\mbox
{\tiny{H}}}^{\,2}}{\raisebox{.3ex}{\mbox{\small{2}}$v$}}
\label{eq:hhh}
\end{equation}
and the self-coupling among four Higgs fields by
\begin{equation}
\lambda_{\mbox{\tiny{HHHH}}} \,\equiv\, \frac{\raisebox{-.3ex}{$\lambda$}}
{\raisebox{.2ex}{\mbox{\small{4}}}} \,=\, \frac{{\rm M}_{\mbox{\tiny{H}}}^
{\,2}}{\mbox{\small{8}}v^{2}}
\end{equation}
Evidently, the Higgs self-couplings are uniquely defined by the mass of the
Higgs boson. The following definitions are often used: $\lambda_{\mbox{\tiny
{HHHH}}} \rightarrow 3!\,\lambda_{\mbox{\tiny{HHHH}}}$ \,and\, $\lambda_{\mbox
{\tiny{HHHH}}} \rightarrow 4!\,\lambda_{\mbox{\tiny{HHHH}}}$.

The covariant derivative in (2) reads
\begin{equation}
D_{\mu} \,=\, i\partial_{\mu} \,+\, g\,{\rm T}_{a\,}W^{a}_{\mu} \,-\,
g'\,\frac{\raisebox{-.3ex}{\rm Y}}{\raisebox{.2ex}{\mbox{\small{2}}}}\,
{\rm B}_{\mu} 
\end{equation}
where ${\rm T}_{a}$ ($a = 1,\,2,\,3$) denote the isospin generators of the 
${\rm SU}(2)_{\mbox{\tiny{L}}}$ gauge group, Y represents the ${\rm U}(1)_
{\mbox{\tiny{Y}}}$ hypercharge generator, $g$ and $g'$ are the electroweak
couplings, and $W^{a}_{\mu}$ and ${\rm B}_{\mu}$ are the gauge fields
associated with the two symmetry groups, respectively. Upon introducing the
physical Higgs field (5) and transforming the electroweak eigenstates $W^{a}
_{\mu}$ and ${\rm B}_{\mu}$ to the mass eigenstates, the kinetic term in (2) 
can be expressed as
\begin{equation}
(D_{\mu}\Phi )^{\dag}(D^{\mu}\Phi ) \,=\, \frac{\raisebox{-.3ex}{\mbox{\small
{1}}}}{\raisebox{.2ex}{\mbox{\small{2}}}}\mbox{\large{$($}}\partial_{\mu\,}
{\rm H}\mbox{\large{$)$}}^{2} \,+\, \frac{g^{2}}{\raisebox{.2ex}{\mbox{\small
{4}}}}\,(v + {\rm H})^{2}\!\left (W^{+}_{\mu\,}W^{-\,\mu} \,+\, \frac{Z_{\mu\,}
Z^{\mu}}{\mbox{\small{2}}\cos^{2\!}\theta_{\mbox{\tiny{W}}}}\right)
\end{equation}
where 
\begin{equation}
\cos\theta_{\mbox{\tiny{W}}} = \frac{g}{\sqrt{g^{2} + g'^{\,2}}},~~~~~~~~~
{\rm e} = g\sin\theta_{\mbox{\tiny{W}}} = g'\cos\theta_{\mbox{\tiny{W}}}
\end{equation}
(e is the electric charge). A comparison with the usual mass terms for the
charged and neutral vector bosons reveals that
\begin{equation}
{\rm M}_{\mbox{\tiny{W}}} \,=\, \frac{gv}{\raisebox{.2ex}{\mbox{\small{2}}}},
~~~~~~~~~{\rm M}_{\mbox{\tiny{Z}}} \,=\, \frac{gv}{2\cos\theta_{\mbox{\tiny
{W}}}} \,=\, \frac{{\rm M}_{\mbox{\tiny{W}}}}{\cos\theta_{\mbox{\tiny{W}}}}
\end{equation}
From (11) we also infer that the Higgs-gauge boson interaction strengths are
\begin{equation}
\lambda_{\mbox{\tiny{HWW}}} \,\equiv\, \frac{g^{2}v}{\raisebox{.2ex}{\mbox
{\small{2}}}} \,=\, \frac{\mbox{\small{2}}{\rm M}_{\mbox{\tiny{W}}}^{\,2}}
{\raisebox{.4ex}{$v$}},~~~~~~~~~\lambda_{\mbox{\tiny{HZZ}}} \,\equiv\, \frac
{g^{2}v}{\mbox{\small{4}}\cos^{2\!}\theta_{\mbox{\tiny{W}}}} \,=\, \frac
{{\rm M}_{\mbox{\tiny{Z}}}^{\,2}}{\raisebox{.4ex}{$v$}}
\end{equation}
and
\begin{equation}
\lambda_{\mbox{\tiny{HHWW}}} \,\equiv\, \frac{g^{2}}{\raisebox{.2ex}{\mbox
{\small{4}}}} \,=\, \frac{{\rm M}_{\mbox{\tiny{W}}}^{\,2}}{v^{2}},~~~~~~~~~
\lambda_{\mbox{\tiny{HHZZ}}} \,\equiv\, \frac{g^{2}}{\mbox{\small{8}}\cos
^{2\!}\theta_{\mbox{\tiny{W}}}} \,=\, \frac{{\rm M}_{\mbox{\tiny{Z}}}^{\,2}}
{\mbox{\small{2}}v^{2}}
\end{equation}

We can relate $v$ to the Fermi constant ${\rm G}_{\mbox{\tiny{F}}} = 1.16639
\times 10^{-5}$ GeV as 
\begin{equation}
\frac{{\rm G}_{\mbox{\tiny{F}}}}{\sqrt{\mbox{\small{2}}}} \,=\, \frac{g^{2}}
{\mbox{\small{8}}{\rm M}_{\mbox{\tiny{W}}}^{\,2}} \,=\, \frac{\raisebox{-.3ex}
{\mbox{\small{1}}}}{\mbox{\small{2}}v^{2}}
\end{equation}
Hence,
\begin{equation}
v \,=\, \mbox{\large{$($}}\sqrt{2}\,{\rm G}_{\mbox{\tiny{F}}}\mbox{\large{$)$}}
^{-1/2} \,\approx\, 246~{\rm GeV}
\end{equation}

The coupling between the Higgs boson and any fermion \mbox{\small{$f$}} is 
given by the interaction Lagrangian
\begin{equation}
{\cal L}_{\mbox{\tiny{H$f\bar{f}$}}} \,=\, -\,\frac{m_{\mbox{\tiny{$f$}}}^{~}}
{\raisebox{.4ex}{$v$}}\,{\rm H}\,\overline{\psi}_{\mbox{\tiny{$f$}}}^{~}\psi_
{\mbox{\tiny{$f$}}}^{~}
\end{equation}
The corresponding coupling strength is 
\begin{equation}
\lambda_{\mbox{\tiny{H$f\bar{f}$}}} \,=\, \frac{m_{\mbox{\tiny{$f$}}}^{~}}
{\raisebox{.4ex}{$v$}}
\end{equation}

The Higgs boson can also couple to two photons. The decay ${\rm H} \rightarrow
\gamma\gamma$ does not occur at lowest level in the Standard Model because
photons couple to charge and the Higgs boson is neutral. The decay proceeds 
through spin-1/2, spin-1 and spin-0 loops. The width is determined by 
\begin{equation}
\Gamma ({\rm H}\rightarrow \gamma\gamma ) \,=\, \frac{\alpha^{2\,}{\rm G}_
{\mbox{\tiny{F}}}}{128\pi^{3}\sqrt{2}}\,{\rm M}_{\mbox{\tiny{H}}}^{\,3}\left |
\sum_{i\,=\,1}^{n}{\rm N}_{ci\,}Q_{i}^{2\,}{\cal F}_{i}\right |^{2}
\end{equation}
where ${\rm N}_{ci}$ is the color multiplicity of particle $i$ (3 for quarks
and 1 otherwise), $Q_{i}$ is the electric charge in units of e, and ${\cal F}_
{i}$ are some functions of $4m_{i}^{2}/{\rm M}_{\mbox{\tiny{H}}}^{\,2}$. The
${\rm H} \rightarrow \gamma\gamma$ decay mode evidently probes the existence of
heavy charged particles. When the particle in the loop is much heavier than the
Higgs boson, ${\cal F}_{0} \rightarrow -1/3$,~~${\cal F}_{1/2}\rightarrow -4/3$
and ${\cal F}_{1} \rightarrow 7$. Note the opposite sign between fermion
and $W$ loops. 

To summarize, Higgs production and decay processes can be computed in the SM
unambiguously in terms of the Higgs mass alone. The Higgs-boson coupling to
fermions and gauge bosons is proportional to the particle masses. We thus
infer that the Higgs boson will be produced in association with heavy
particles, and will decay into the heaviest particles that are kinematically
accessible.

The discovery of a Higgs boson with a mass below about 135~GeV might
indicate that the Standard Model is embedded in a {\small\bf
supersymmetric theory}. The minimal supersymmetric extension of the
Standard Model (MSSM) introduces two SU(2) doublets of complex Higgs
fields, whose neutral components have vacuum expectation values
$v_{\mbox{\tiny{1}}}$ and $v_{\mbox{\tiny{2}}}$. In this model,
spontaneous electroweak symmetry breaking results in five physical
Higgs-boson states: two neutral scalar fields $h^0$ and H$^0$, a
pseudoscalar $A^0$ and two charged bosons H$^\pm$. This extended Higgs
system can be described at `tree level' by two parameters: the ratio
$\tan\beta \equiv v_{\mbox{\tiny{2}}}/v_{\mbox{\tiny{1}}}$, and a mass
parameter, which is generally identified with the mass of the
pseudoscalar boson $A^0$, \MA. While there is a bound of about 135~GeV
on the mass of the lightest $CP$-even neutral Higgs boson $h^0$
\cite{h0-mass1,h0-mass2}, the masses of the H$^0$, $A^0$ and H$^\pm$
bosons may be as large as 1~TeV.

The trilinear self-coupling of the lightest MSSM Higgs boson at `tree level' is
given by
\begin{equation}
\lambda_{hhh}  \,=\, \frac{{\rm M}_{\mbox{\tiny{Z}}}^{\,2}}{\raisebox{.3ex}
{\mbox{\small{2}}$v$}}\cos 2\alpha\sin(\beta+\alpha),
\end{equation}
where
\begin{equation}
\tan 2\alpha\,=\, \tan 2\beta \,
\frac{{\rm M}_{\mbox{\tiny{A}}}^{\,2}+{\rm M}_{\mbox{\tiny{Z}}}^{\,2}}
{{\rm M}_{\mbox{\tiny{A}}}^{\,2}-{\rm M}_{\mbox{\tiny{Z}}}^{\,2}}.
\end{equation}
We see that for arbitrary values of the MSSM input parameters $\tan\beta$ and
\MA, the value of the $h^0$ self-coupling differs from that of the SM Higgs
boson. However, in the so-called `decoupling limit' ${\rm M}_{\mbox{\tiny{A}}}
^{\,2}\sim {\rm M}_{\mbox{\tiny{H}}^{0}}^{\,2} \sim {\rm M}_{\mbox{\tiny
{H}}^{\pm}}^{\,2}\gg v^2/2$, the trilinear and quadrilinear self-couplings of
the lightest $CP$-even neutral Higgs boson $h^0$ approach the SM value. The
inclusion of one-loop MSSM Higgs-sector corrections and ${\cal O}(m_t^4)$ 
Yukawa corrections does not lead to any significant deviations from the SM 
prediction \cite{h0-self-coupling-MSSM}. As a result, the $h^0$-boson 
self-interactions in the `decoupling limit' do not differ appreciably from
those of the SM Higgs particle. 

In the non-supersymmetric two-Higgs-doublet model (the simplest extension of
the SM), large one-loop effects can occur. For charged Higgs bosons with masses
of about 400~GeV, the decay widths of $h^0\to\gamma\gamma$, $h^0\to\gamma Z$
and $h^0\to b\bar b$ may differ from the SM values by as much as $10\%-25\%$
\cite{Arhrib:ak}. In this model, the non-decoupling effects of the additional 
heavier Higgs bosons in loops can produce ${\cal O}(100\%)$ deviations of the
effective $h^0h^0h^0$ self-coupling from the SM prediction, even if the Higgs
couplings to gauge bosons and fermins are almost SM-like 
\cite{Kanemura:2002vm}.

The precision electroweak data obtained over the past sixteen years consists of
over a thousand individual measurements. Many of these measurements may be
combined to provide a global test of consistency with the Standard Model. The
best constraint on ${\rm M}_{\mbox{\tiny{H}}}$ is obtained by making a global
fit to the data, which yields ${\rm M}_{\mbox{\tiny{H}}} = 91^{\,+\,58}_
{\,-\,37}$ GeV \cite{grunewald}. The precision electroweak data, therefore,
strongly suggest that the most likely mass for the SM Higgs boson is just above
the limit of 114.4 GeV set by direct searches at the LEP $e^{+}e^{-}$ collider
\cite{LEP}.

The next crucial step in our investigation of the Standard Model would be to
discover the Higgs boson and determine its properties. Ideally, one would like
to determine, {\em in a model-independent way}, the mass, total width, spin,
parity and CP properties of the Higgs boson, as well as its tree-level and
one-loop induced couplings. 
In contrast to any anomalous couplings of the gauge bosons, an
anomalous self-coupling of the Higgs particle would contribute to
electroweak observables only at two-loop and higher orders, and is
therefore virtually unconstrained by the current precision
measurements \cite{vanderBij:1985ww}.

\vspace*{0.8cm} 
\section{Raisons d'\^{e}tre for a photon collider}
\vspace*{0.8cm}

~~~~Once the Higgs boson has been discovered, a thorough exploration of the
Higgs sector of the Standard Model will be undertaken with hadron, $e^{+}e^{-}$
and photon colliders. The rich set of final states in $\gamma\gamma$, $pp$ and
$e^{+}e^{-}$ collisions will play an essential role in measuring the mass,
two-photon width, spin and parity of the Higgs boson, which are difficult to
determine with only one initial state. By combining data from $e^{+}e^{-}$ 
and $\gamma\gamma$ collisions, the total decay width of the Higgs boson can be
determined in a model-independent way with a precision of about 10\% (see
\cite{belusev} and references therein). 

Since photons couple directly to all fundamental fields carrying the 
electromagnetic current (leptons, quarks, $W$ bosons, supersymmetric particles),
$\gamma\gamma$ collisions provide a comprehensive means of exploring virtually
every aspect of the SM and its extensions. The production mechanisms in $e^{+} 
e^{-}$ collisions are often more complex and model-dependent. In $\gamma\gamma$
collisions, the Higgs boson will be produced as a single resonance in a state
of definite CP, which is perhaps the most important advantage over $e^{+}e^{-}$
annihilations, where this $s$-channel process is highly suppressed. For the
Higgs-boson mass in the range 115$-$200 GeV, the effective cross-section for
$\gamma\gamma \rightarrow {\rm H}$ is about an order of magnitude larger than
that for Higgs production in $e^{+}e^{-}$ annihilations. In this mass range, 
the process $e^{+}e^{-} \rightarrow Z{\rm H}$ requires considerably higher 
centre-of-mass (CM) energies than $\gamma\gamma \rightarrow {\rm H}$. Since 
$\gamma\gamma\rightarrow{\rm H}$ proceeds through a `loop diagram' and receives
contributions from {\em all} particles with mass and charge, this mode is a
powerful probe of new physics beyond the SM. Moreover, we find that the 
sensitivity of the cross-section $\mbox{\large{$\sigma$}}_{\gamma\gamma\,
\rightarrow\,\mbox{\tiny{HH}}}$ to the trilinear Higgs self-coupling is maximal
near the $2{\rm M}_{\mbox{\tiny{H}}}$ threshold for ${\rm M}_{\mbox{\tiny{H}}}
= 115-150$ GeV, and is {\em larger} than the sensitivities of $\mbox{\large
{$\sigma$}}_{e^{+}e^{-}\,\rightarrow\,\mbox{\tiny{$Z$HH}}}$ and $\mbox{\large
{$\sigma$}}_{e^{+}e^{-}\,\rightarrow\,\nu\bar{\nu}\mbox{\tiny{HH}}}$ to this
coupling for $2{\rm E}_{e} \lsim 700$ GeV.

\vspace*{0.8cm}
\section{Higgs-pair production in $\gamma\gamma$ and $e^+e^-$ collisions}
\vspace*{0.8cm}

\begin{figure}[htb]
\setlength{\unitlength}{1cm}
\begin{picture}(10,10)
\put(0.25,0){\epsfig{file=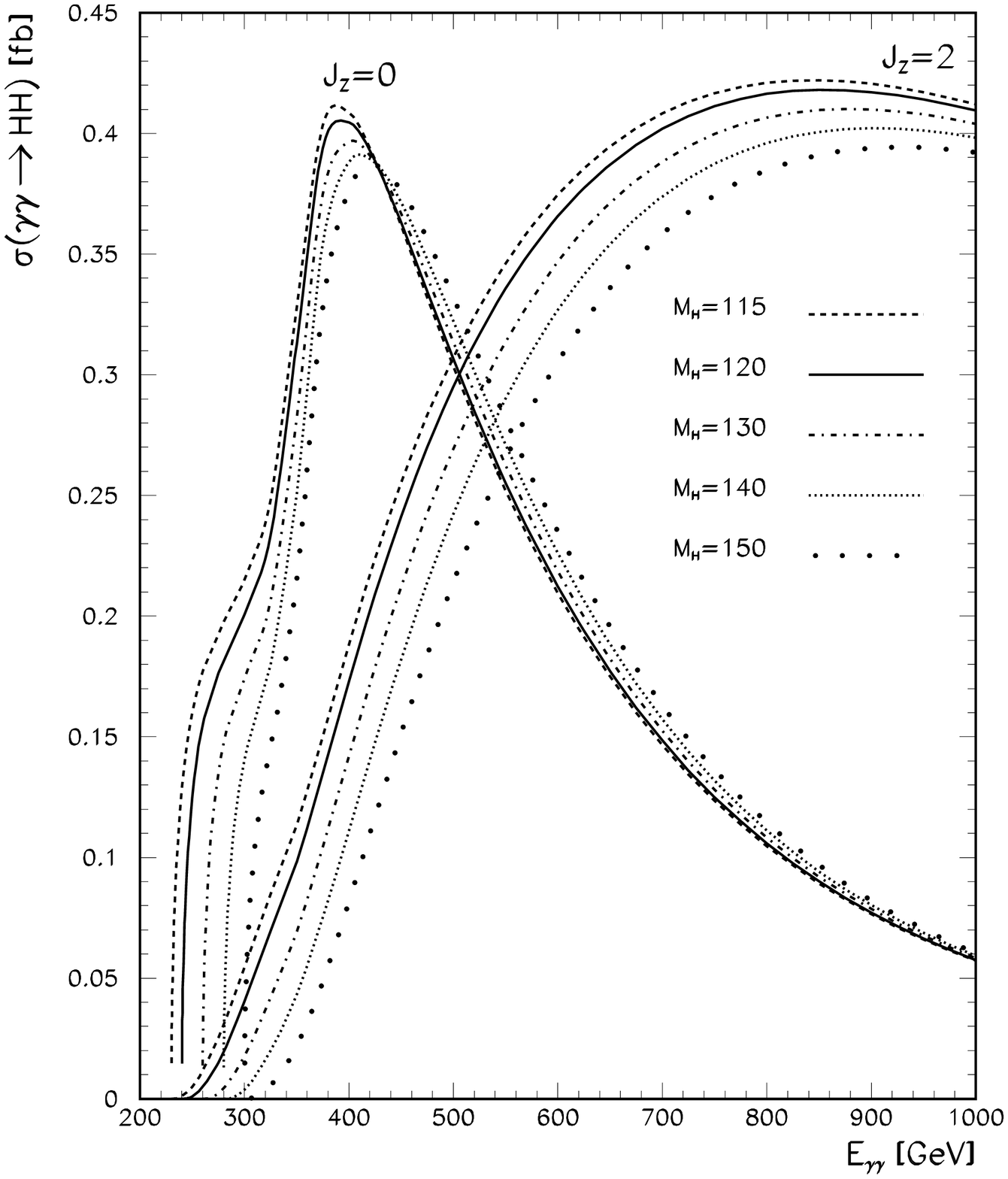,width=0.5\textwidth}}
\put(8.5,0){\epsfig{file=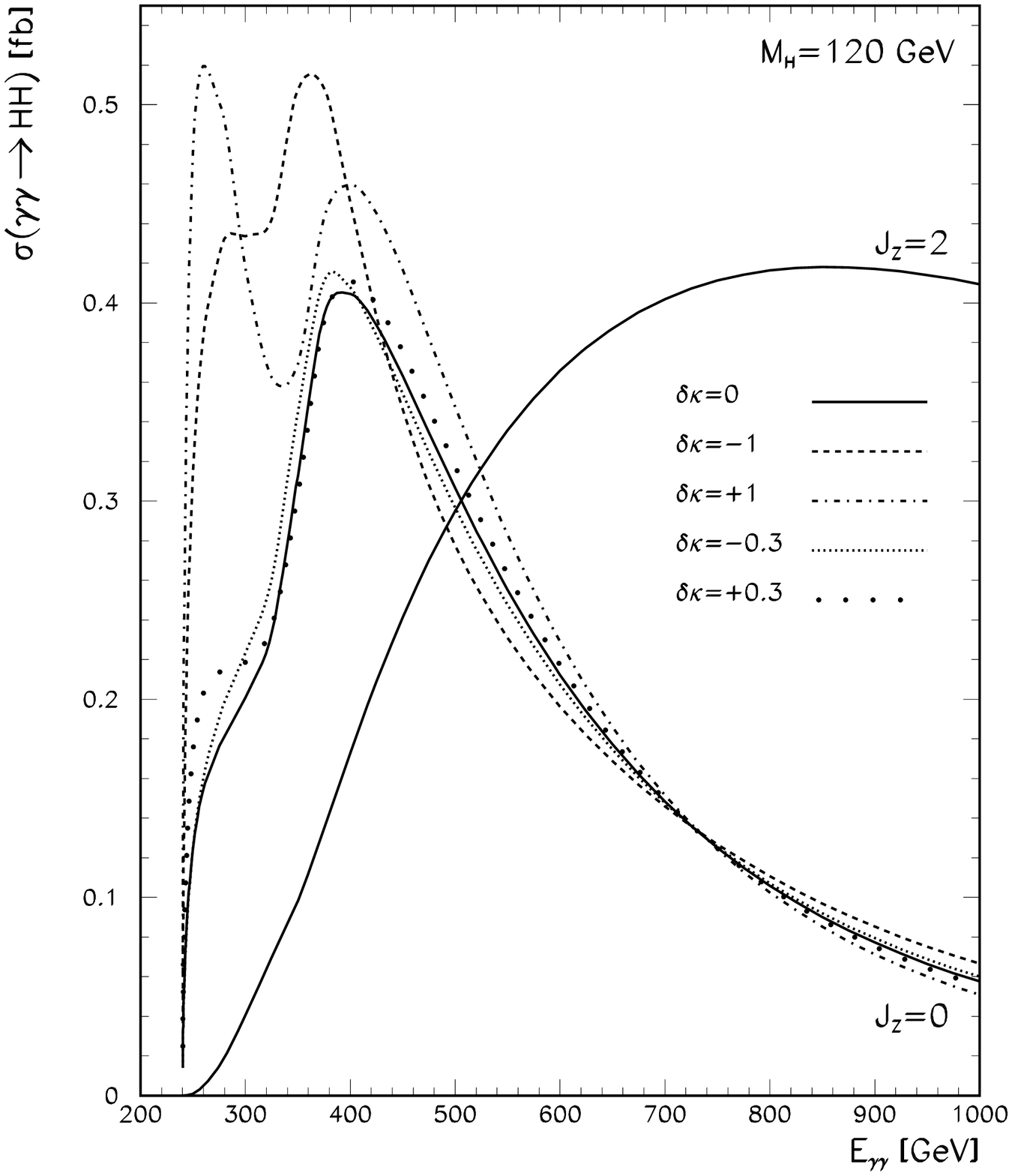,width=0.5\textwidth}}
\end{picture}
\caption{(a) The total $\gamma\gamma\to{\rm HH}$ cross-section as a function of
  the $\gamma\gamma$ centre-of-mass energy for \MH=115, 120, 130, 140
  and 150~GeV. Contributions for equal ($J_z=0$) and opposite
  ($J_z=2$) photon helicities are shown separately.
  \newline 
  (b) The cross-sections for HH production in $\gamma\gamma$ collisions for
anomalous trilinear Higgs self-couplings $\delta\kappa=0,\pm 1, \pm 0.3$.} 
\label{fig:gghh}
\end{figure}

~~~~The production of a pair of SM Higgs bosons in  photon-photon collisions
\begin{equation}
\gamma\gamma \,\to\, {\rm HH} 
\label{eq:gghh}
\end{equation}
which is related to the Higgs-boson decay into two photons, is due to $W$-boson
and top-quark box and triangle loop diagrams. The total cross-section for
$\gamma\gamma\to{\rm HH}$ in polarized photon-photon collisions, calculated at
the leading one-loop order \cite{Jikia:1992mt} as a function of the $\gamma
\gamma$ centre-of-mass energy and for ${\rm M}_{\mbox{\tiny{H}}} = 115-150$ 
GeV, is shown in Fig.\,\ref{fig:gghh}a. The cross-sections calculated for equal
($J_z=0$) photon helicities, $\mbox{\large{$\sigma$}}_{\gamma\gamma\,
\rightarrow\,\mbox{\tiny{HH}}}(\mbox{\small{$J_{z}=0$}})$, and for different 
values of ${\rm M}_{\mbox{\tiny{H}}}$ rise sharply above the HH-threshold, and
each has a peak value of about $0.4$~fb at a $\gamma\gamma$ centre-of-mass
energy of 400~GeV. In contrast, the cross-sections $\mbox{\large{$\sigma$}}_
{\gamma\gamma\,\rightarrow\,\mbox{\tiny{HH}}}(\mbox{\small{$J_{z}=2$}})$ rise
much slower with energy, because a pair of Higgs bosons is produced in a state
with orbital angular momentum of at least 2; each of these cross-sections
reaches a value of about $0.4$~fb at a $\gamma\gamma$ centre-of-mass energy
of 800~GeV. 

The cross-sections for equal photon helicities are of special interest, since
only the $J_z=0$ amplitudes contain contributions with trilinear Higgs 
self-coupling. By adding to the SM Higgs potential (\ref{eq:potential}) a 
gauge-invariant dimension-6 operator $\mbox{\large{$($}}\Phi^{\dag}\Phi\mbox
{\large{$)$}}^{3}$, one can introduce a gauge-invariant anomalous trilinear
Higgs coupling $\delta\kappa$ \cite{Jikia:1992mt}. For the reaction 
(\ref{eq:gghh}), the only effect of such a coupling in the {\em unitary gauge}
would be to replace the trilinear HHH coupling of the SM, Eq. (\ref{eq:hhh}),
by an anomalous Higgs self-coupling 
\begin{equation}
\widetilde{\lambda}_{\mbox{\tiny{HHH}}}\, = \, 
(1+\delta\kappa)\lambda_{\mbox{\tiny{HHH}}}
\end{equation}
The dimensionless anomalous coupling $\delta\kappa$ is normalized in such a way
that $\delta\kappa=-1$ exactly cancels the SM HHH coupling. The cross-sections
$\mbox{\large{$\sigma$}}_{\gamma\gamma\,\rightarrow\,\mbox{\tiny{HH}}}$ for
various values of $\delta\kappa$ are shown in Fig.~\ref{fig:gghh}b. 

In an experiment to measure the trilinear Higgs self-coupling, the contribution
from $\gamma\gamma \,\to\, {\rm HH}$ for opposite photon helicities represents
an irreducible background. Clearly, the optimal energy for such measurements
would be somewhere between the production threshold and 400~GeV. In order to 
ascertain the potential of $\gamma\gamma$ colliders for measuring an anomalous
Higgs self-coupling, one must take into account the fact that photon spectra
will not be monochromatic \cite{Ginzburg:1982yr}. The cross-section for 
Higgs-pair production in polarized $\gamma\gamma$ collisions is given by
\begin{equation}
\mbox{\large{$\sigma$}}_{\gamma\gamma\,\rightarrow\,\mbox{\tiny{HH}}}^{~} =
\int_{4{\rm M}_{\mbox{\tiny{H}}}^{\,2}/s}^{y_m^2} d\tau
\frac{d L_{\gamma\gamma}}{d\tau}\!
\left[\frac{1}{2}\!\left(1+\langle\xi_2^{(1)}\xi_2^{(2)}\rangle\right)\!
\hat\sigma_{++}(\hat s)
 + \frac{1}{2}\!\left(1-\langle\xi_2^{(1)}\xi_2^{(2)}\rangle\right)\!
\hat\sigma_{+-}(\hat s)\right]
\end{equation}
where
\begin{eqnarray}
&&\frac{d L_{\gamma\gamma}}{d\tau} = \int_{\tau/y_m}^{y_m}\frac{dy}
{\raisebox{.5ex}{$y$}}f_\gamma(x,y)f_\gamma(x,\tau /y), 
\nonumber\\*[2mm]
&&\tau=\frac{\hat s}{\raisebox{.5ex}{$s$}},\quad 0\,\leq\, y=\frac{E_\gamma}
{E_e}\, \leq \, y_m=\frac{x}{x+1},\quad x\equiv\frac{4E_e\omega_0}{m_e^2}.
\end{eqnarray}
Here $E_e$ is the energy of the electron beam, $\omega_0$ is the laser
photon energy, $f_\gamma(x,y)$ is the photon momentum distribution
function and $\xi^{(1,2)}_2$ are mean photon helicities
\cite{Ginzburg:1982yr};  $\hat\sigma_{++}$ and $\hat\sigma_{+-}$ are the
cross-sections for Higgs-pair production, calculated assuming  
monochromatic photons with total helicities $J_z=0$ and
$J_z=2$, respectively. As usual, the dimensionless parameter $x$ has been set
to 4.8 ($y_m\approx0.8$) to avoid undesirable backgrounds. In what follows we
shall assume 90\% polarization for electron beams and 100\% for laser beams, in
a configuration that maximizes the $\gamma\gamma$ luminosity for $J_z=0$ in the
high-energy part of the photon spectrum \cite{Telnov:2003ua}: 
\begin{equation}
L_{\gamma\gamma} = \int_{\tau=(0.8\,y_m)^2}^{y_m^2} 
d\tau\,\frac{d L_{\gamma\gamma}}{d\tau} \,\approx\, (1/3)L_{e^+e^-}.
\label{eq:Lgg}
\end{equation}

\begin{figure}[htb]
\begin{center}
\epsfig{file=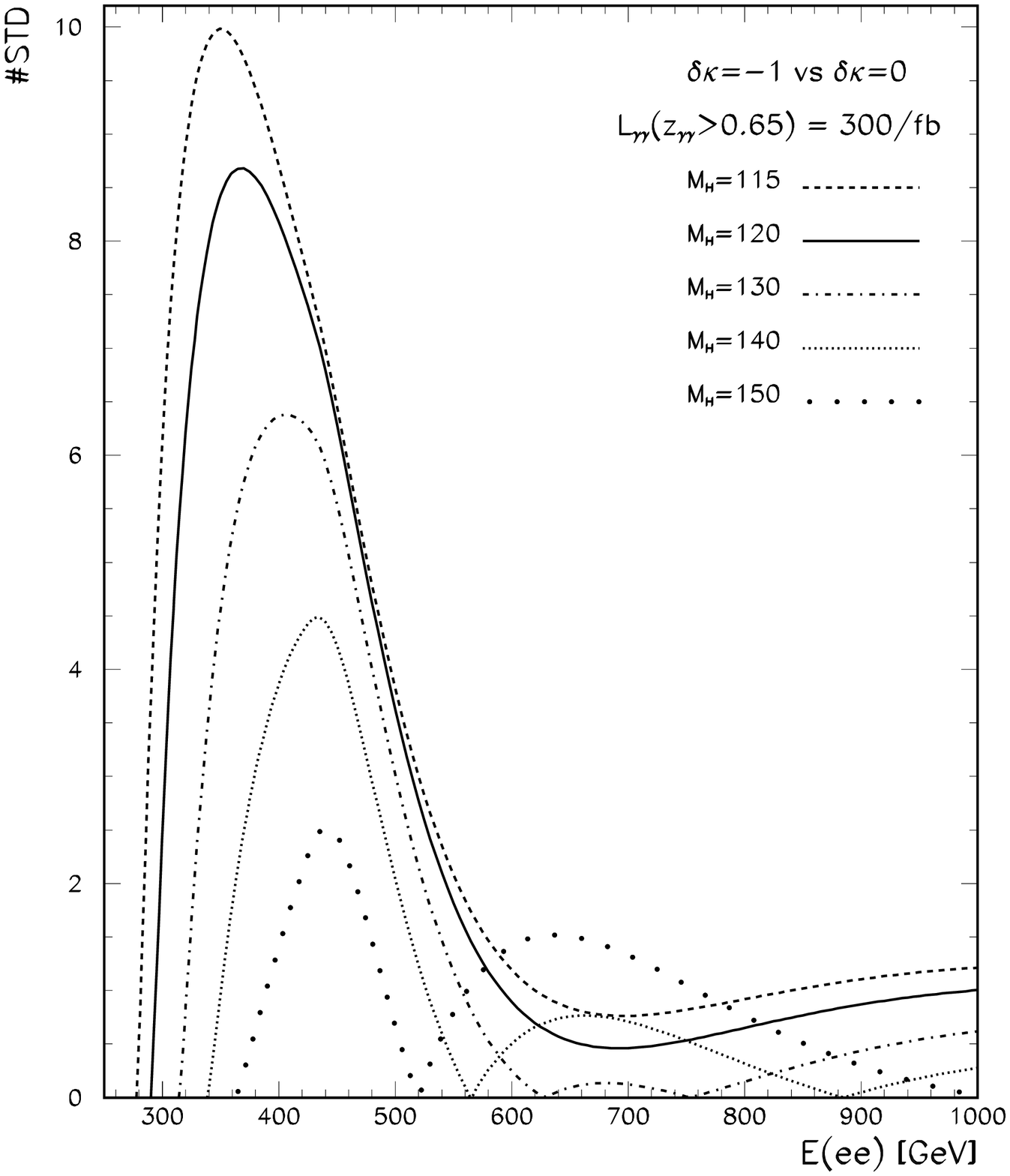,height=0.45\textheight}
\end{center}
\caption{For the process $\gamma\gamma\to{\rm HH}$, the number of standard 
  deviations from the SM prediction for event rates, defined by Eq.
  (\ref{eq:std_gg}), is plotted as a function of the $e^-e^-$ centre-of-mass
  energy assuming a $\gamma\gamma$ luminosity 
  $L_{\gamma\gamma}=300$~fb$^{-1}$.} 
\label{fig:std_gg}
\end{figure}

In terms of standard deviations, the discrepancy between the SM prediction for
event rates and that for zero HHH coupling is defined by
\begin{equation}
\mbox{\#STD} \,=\, \frac{|\mbox{\large{$\sigma$}}(\delta\kappa=0) - \mbox
{\large{$\sigma$}}(\delta\kappa=-1)|}{\sqrt{\mbox{\large{$\sigma$}}(\delta
\kappa=0)}}\sqrt{L_{\gamma\gamma}}
\label{eq:std_gg}
\end{equation}
(see Fig.~\ref{fig:std_gg}). Here the cross-sections are given by Eq. (25), an
efficiency of 100\% is assumed and the $\gamma\gamma$ luminosity is taken to
be 300~fb$^{-1}$ (see (\ref{eq:Lgg})). For \MH=120~GeV, a maximum sensitivity
of almost 9$\sigma$ is achieved at a $e^-e^-$ centre-of-mass energy of 370~GeV.
An effect of more than 5$\sigma$ is seen already at energies above 310~GeV.
Note that the abscissa in Fig.~\ref{fig:std_gg} shows the $e^-e^-$ CM energies.
For instance, $E_{e^-e^-}=310$~GeV corresponds to a maximum $\gamma\gamma$ CM 
energy of $y_m E_{e^-e^-}\approx 254$~GeV, which is just 14~GeV above the
HH-threshold. Of course, the numbers shown in Fig.~\ref{fig:std_gg} represent
the maximum achievable sensitivity assuming 100\% detection and reconstruction
efficiencies and no backgrounds. In reality, the sensitivity to an anomalous 
Higgs self-coupling will be considerably worse. Nevertheless, 
Fig.~\ref{fig:std_gg} shows that in photon-photon collisions the optimum
$e^-e^-$ CM energy for measuring the trilinear Higgs self-coupling is rather
low (between 300 and 400~GeV) for a Higgs-boson mass \MH$\leq 130$~GeV.

It is well known that hadron colliders are not well suited for measuring the
self-coupling of the Higgs boson if \MH$\leq 140$~GeV \cite{Baur:2003gp}. The
potential of a future $e^+e^-$ collider for determining the HHH coupling has 
therefore been closely examined 
\cite{Djouadi:1999gv,Miller:1999ct,Castanier:2001sf,Yasui:2002se,
Belanger:2003ya}. The trilinear Higgs-boson self-coupling can be
measured either in the double Higgs-strahlung process
\begin{equation}
e^+e^-\, \to \, Z{\rm HH} 
\label{eq:eezhh}
\end{equation}
or in the $W$-boson fusion reaction
\begin{equation}
e^+e^-\, \to \, \nu_e\bar\nu_e{\rm HH}.
\label{eq:eennhh}
\end{equation}

\begin{figure}[htb]
\setlength{\unitlength}{1cm}
\begin{picture}(10,10)
\put(0.25,0){\epsfig{file=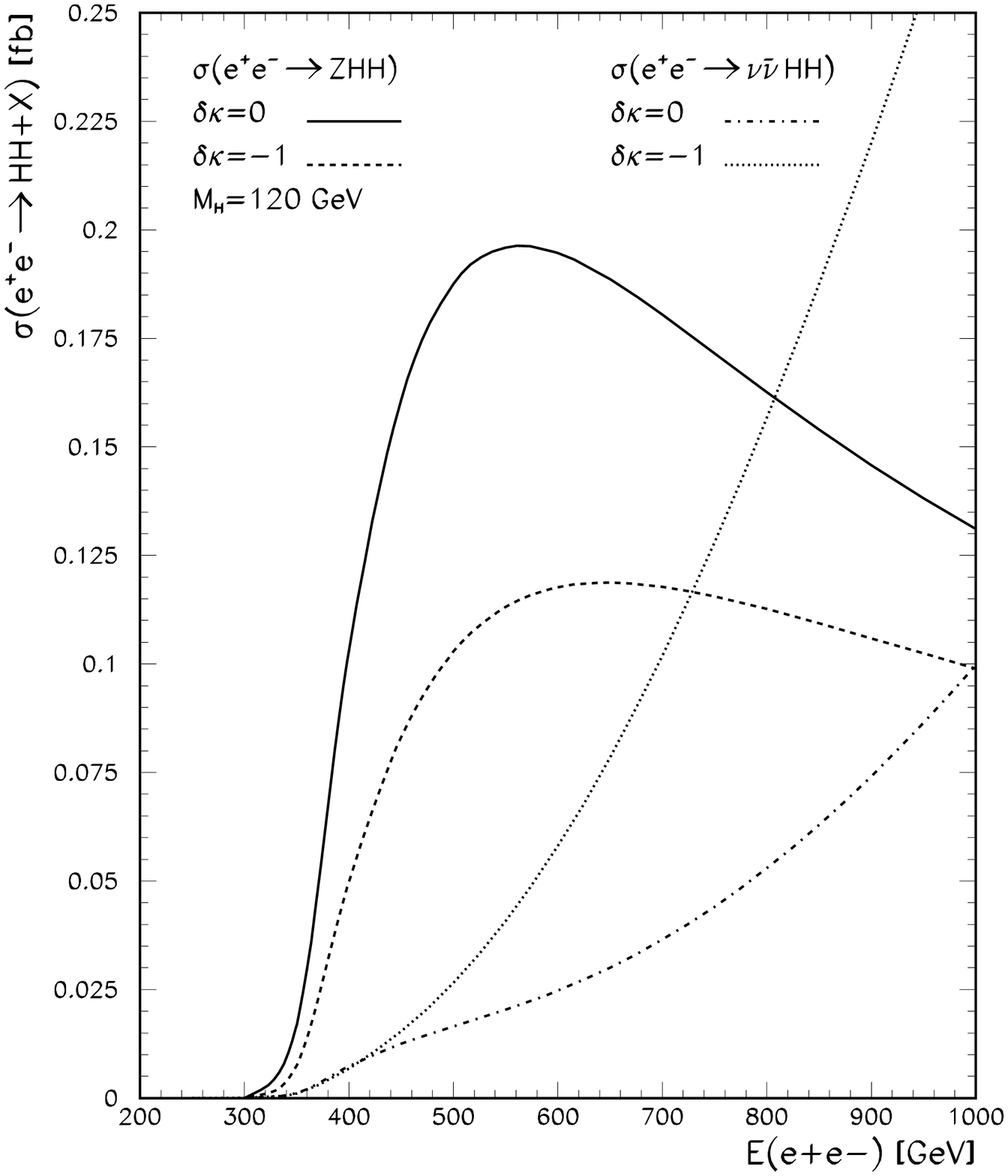,width=0.5\textwidth}}
\put(8.5,0){\epsfig{file=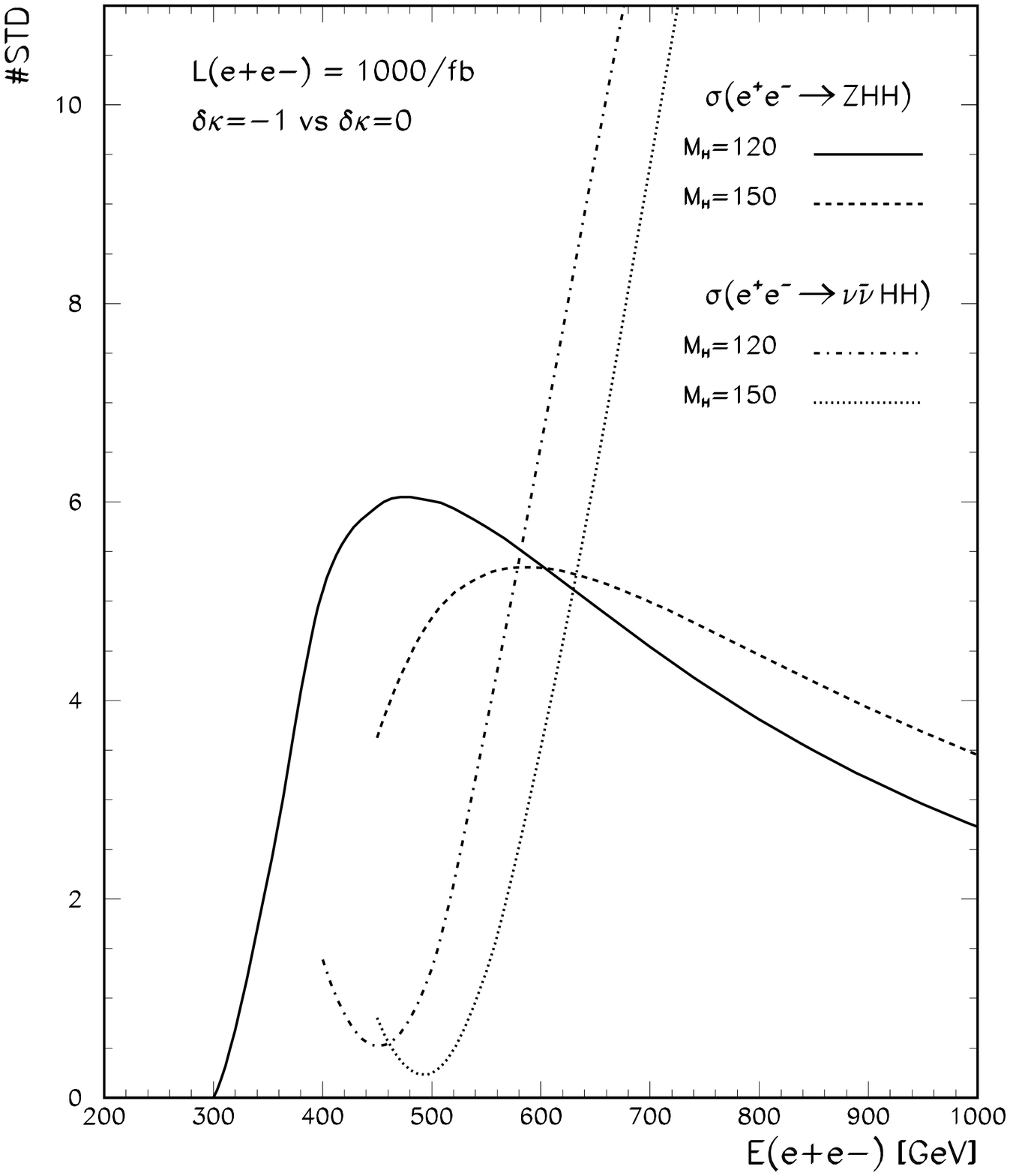,width=0.5\textwidth}}
\end{picture}
\caption{(a) The total cross-sections for $e^+e^-\, \to \, Z{\rm HH}$ and 
  $e^+e^-\, \to \,\nu_e\bar\nu_e{\rm HH}$ as functions of the $e^+e^-$
  centre-of-mass energy for \MH=120~GeV and anomalous trilinear
  Higgs self-couplings $\delta\kappa=0,-1$.
  \newline 
  (b) For HH production in $e^+e^-$ collisions, the number of standard
  deviations from the SM prediction for event rates, defined analogously to 
  Eq. (\ref{eq:std_gg}), is plotted as a function of the $e^+e^-$
  centre-of-mass energy assuming an $e^+e^-$ luminosity  
  $L_{e^+e^-}=1000$~fb$^{-1}$.} 
\label{fig:eehh}
\end{figure}

The total cross-section for the Higgs-pair production in $e^+e^-$
collisions, calculated for {\em unpolarized} beams, is presented in
Fig~\ref{fig:eehh}a. If the electron beams are 100\% polarized, the 
cross-section for the reaction (\ref{eq:eezhh}) will approximately stay the
same, but the cross-section for the $W$-fusion process (\ref{eq:eennhh})
will be twice as large. The cross-sections shown in Fig~\ref{fig:eehh}a were 
calculated at `tree level' using the program CompHEP \cite{Pukhov:1999gg}.
The effect of full ${\cal O}(\alpha)$ electroweak radiative corrections to
the process (\ref{eq:eezhh}) has been shown to be small around the peak of the
corresponding cross-section \cite{Belanger:2003ya}. From Fig~\ref{fig:eehh}a we
infer that the SM cross-section for the process (\ref{eq:eezhh}) exceeds 0.1~fb
at 400~GeV for \MH=120~GeV, and reaches a broad maximum of about 0.2~fb at a
$e^+e^-$ centre-of-mass energy of 550~GeV. The SM cross-section for the 
$W$-boson fusion process (\ref{eq:eennhh}) stays below 0.1~fb all the way up to
$E_{e^{+}e^{-}} \approx 1$ TeV.

The cross-sections for the processes (\ref{eq:eezhh}) and (\ref{eq:eennhh}),
computed for $\delta\kappa=0,-1$ and \MH=120~GeV, are shown in 
Fig.~\ref{fig:eehh}a. The number of standard deviations from the SM prediction
for event rates, defined analogously to Eq. (\ref{eq:std_gg}), is shown in 
Fig.~\ref{fig:eehh}b for \MH=120 and 150~GeV. We again assume an efficiency of
100\%, but the $e^+e^-$ luminosity is taken to be 1000~fb$^{-1}$. The 
polarization of the electron beams was chosen to be 90\%. 
For \MH=120~GeV, a maximum sensitivity of about 6$\sigma$ is achieved
at a centre-of-mass energy of 500~GeV in the double Higgs-strahlung
process (\ref{eq:eezhh}). An effect of more than 5$\sigma$ is seen at
energies above 400~GeV.  A comparison of Fig.~\ref{fig:std_gg} and
Fig.~\ref{fig:eehh}b reveals that the optimum $e^+e^-$ centre-of-mass
energy for measuring the Higgs self-coupling in the reaction
(\ref{eq:eezhh}) is about 500~GeV, significantly higher than the
corresponding energy in $\gamma\gamma$ collisions. 

\vspace*{0.8cm}
\section{Backgrounds}
\vspace*{0.8cm}

~~~~We shall present an order-of-magnitude estimate of the most important 
backgrounds to the process $\gamma\gamma\to{\rm HH}$. The dominant background
is the $W$-boson pair production $\gamma\gamma\to W^+W^-$, with the total
cross-section of 70~pb at 300~GeV. However, by imposing the invariant mass cut
\begin{equation}
|{\rm M}(q\bar q) - {\rm M}_{\mbox{\tiny{H}}}| < 5\mbox{~GeV}
\label{eq:masscut}
\end{equation}
the $W^{+}W^{-}$ background can be reduced by about four orders of magnitude. 
In order to suppress this background even further, one could rely on the fact 
that the predominant decay mode of the SM Higgs boson with ${\rm M}_{\mbox
{\tiny{H}}} = 115$ to 130~GeV is into a pair of $b$ quarks, with a SM branching
ratio that decreases from 73\% to 53\% as the Higgs-boson mass increses.
This is the dominant decay mode also of the MSSM $h^0$ boson for various values
the MSSM parameters, in particular for $\tan\beta>1$.
In order to select the ${\rm HH}\to b\bar b b\bar b$ events, we require that
at least three jets be identified as originating from $b$-quarks. If we assume
that the standard method used for tagging $b$-hadrons at the LEP $e^{+}e^{-}$
collider \cite{Abdallah:2002xm} can also be used at a photon collider,
then the sample tagged as $b$-quark would have the following flavour 
composition: 4.3\% light quarks, 10.4\% $c$-quarks and 85.4\% $b$-quarks 
\cite{Abdallah:2002xm}. The requirement that at least three jets originating 
from $W^\pm$ decays be identified as $b$-jets would suppress
$\gamma\gamma\to W^+W^-$ by another three orders of magnitude, to a 
level well below the HH signal.

\begin{table}[ht]
\begin{center}
\caption{
The cross-sections for the production of four heavy quarks in unpolarized
$\gamma\gamma$ collisions for $E_{\gamma\gamma}=250$ and 300~GeV. }
\vspace*{0.2cm}
\begin{tabular}{|c|c|c|c|}\hline
\rule[0mm]{0mm}{4.5mm}
\Egg= 250 (300)~GeV &
$\mbox{\large{$\sigma$}}_{\mbox{\scriptsize{tot}}}^{~}$ & 
$|\cos\theta_{q,\bar{q}}|<0.9$ &
$|\cos\theta_{q,\bar{q}}|<0.9$\\
\rule[-2mm]{0mm}{7mm}
\MH = 120 GeV & 
(\mbox{\small{fb}}) & & 115~GeV $< {\rm M}_{q\bar{q}}<$125~GeV\\
\hline
\rule[0mm]{0mm}{4.5mm} 
$\gamma\gamma\to b\bar{b}b\bar{b}$ &
360 (380) &  5.0 (3.9) & 0.015 (0.015)\\ 
\hline 
\rule[0mm]{0mm}{4.5mm} 
$\gamma\gamma\to b\bar{b}c\bar{c} $ &
9400 (9800) &  66 (52) & 0.13 (0.16)\\
\hline 
\rule[0mm]{0mm}{4mm} 
$\gamma\gamma\to c\bar{c}c\bar{c} $ &
81000 (83000) & 150 (120) & 0.24 (0.26)\\
\hline 
\end{tabular}
\end{center}
\end{table}

The next most significant background is the direct
production of four heavy quarks in photon-photon collisions. The total
cross-sections for $b\bar{b}b\bar{b}$, $b\bar{b}c\bar{c}$ and
$c\bar{c}c\bar{c}$ production are shown in Table~1. These
cross-sections do not decrease with energy and are quite large. For instance,
the cross-section for the production of four $c$-quarks is even larger
than that for $W^+W^-$ production. Since the cross-sections calculated 
for two-photon helicities $J_z=0$ and $J_z=2$ have similar magnitudes,
the polarization of photon beams does not lead to a reduction in the
four-quark background. The $b$ and $c$ quarks are produced mostly in 
the forward or backward direction. As shown in 
Table~1, a simple angular cut
\begin{equation}
|\cos\theta_{q,\bar{q}}|<0.9
\label{eq:anglecut}
\end{equation}
suppresses these backgrounds by at least two orders of magnitude.
Near the HH production threshold, the angular distribution of $b$-jets
originating from Higgs-boson decays is isotropic, and the
efficiency of the angular cut (\ref{eq:anglecut}) is about 80\%.
Since the cross-sections for quark production are still much larger than the 
cross-section for double Higgs-boson production after the cut
(\ref{eq:anglecut}), the invariant-mass cut (\ref{eq:masscut}) should also
be imposed. As shown in Table~1, after these two cuts the
cross-section for $b\bar{b}b\bar{b}$ production is already an order of 
magnitude smaller than $\mbox{\large{$\sigma$}}_{\gamma\gamma\,\rightarrow\,
\mbox{\tiny{HH}}}$, and the cross-sections for $b\bar{b}c\bar{c}$ and
$c\bar{c}c\bar{c}$ production are of the same order as $\mbox{\large{$\sigma$}}
_{\gamma\gamma\,\rightarrow\,\mbox{\tiny{HH}}}$. 
The additional requirement that at least three jets be identified as
$b$-jets would suppress these cross-sections well below that of the signal. 
After the invariant-mass cut (\ref{eq:masscut}), the angular cut
(\ref{eq:anglecut}) and the $b$-tagging requirement, the reconstruction
efficiency for the HH final state is about 50\%. A more thorough study
will definitely improve this number.

Other potential background sources are $\gamma\gamma\to b\bar b Z$,
$\gamma\gamma\to c\bar c Z$, $\gamma\gamma\to q\bar{q}^{\,\prime\,} W$,
$\gamma\gamma\to W^+W^- Z$ and $\gamma\gamma\to ZZ$ processes. We 
believe that appropriate invariant-mass and angular cuts, as well as the
$b$-jet tagging requirement, would suppress these backgrounds to a manageable
level. 

\vspace*{0.8cm}
\section{The proposed facility}
\vspace*{0.8cm}

~~~~We propose the construction of an X-band $e^{-}e^{-}$ linac (based on the
JLC design) and a terawatt laser system (based on the {\em Mercury}
architecture) in order to produce Compton-scattered $\gamma$-ray beams for a 
photon collider \cite{belusev}. The key advantage of using $e^{-}e^{-}$ beams 
is that they can be polarized to a high degree (about 90\%).\,\footnote{~Both
the energy spectrum and polarization of the backscattered photons depend 
strongly on the polarizations of the incident electrons and laser photons. By
polarizing the incident beams one can tailor the photon energy distribution to
one's needs \cite{telnov}. In a collision of two photons, the possible
helicities are 0 or 2. For example, the Higgs boson is produced in the $J_{z} =
0$ state, whereas the background processes $\gamma\gamma \rightarrow b\bar{b},
\,c\bar{c}$ are suppressed for this helicity configuration. The circular
polarization of the photon beams is therefore an important asset, for it can be
used both to enhance the signal and suppress the background.} In $\gamma\gamma$
collisions, a light Higgs boson can be detected either as a peak in the  
invariant mass distribution or by conducting an energy scan exploiting the 
sharp high-energy edge of the $\gamma\gamma$ luminosity distribution
\cite{belusev}. The proposed facility would use 40-MeV/m rf cavities in a 7-km
tunnel to reach a centre-of-mass energy $2{\rm E}_{e} = 200$ GeV (${\rm E}_
{\gamma\gamma} \approx 160$ GeV). It would be capable of producing around
$10^{4}$ light Higgs bosons per year. 

We envisage to add a positron source to the linac, turning it into a 
high-luminosity $e^{+}e^{-}$ collider \cite{belusev1}. Such a machine would 
operate in a wide energy range, from the $Z^{0}$ peak to well above the $WW$
threshold. High-precision studies of electroweak physics provide a natural 
complement to the direct searches for the Higgs boson. In principle, all
measurements done at LEP and SLC could be repeated at the proposed $e^{+}e^{-}$
collider with much better accuracy. Assuming a geometric luminosity ${\cal L}_
{e^{+}e^{-}} \approx 5\times 10^{33}~{\rm cm}^{-2}\,{\rm s}^{-1}$ at the 
$Z^{0}$ resonace, and the cross-section $\mbox{\large{$\sigma$}}_{\mbox{\tiny
{$Z$}}}^{~} \approx 30$ nb, about $2\times 10^{9}$ $Z^{0}$ events would be
produced in an operational year of $10^{7}$ s, which is approximately 200 times
the entire LEP statistics. Moreover, about $10^{6}$ $W$ bosons could be 
detected near the $W$-pair threshold at the optimal energy point for measuring
the $W$-boson mass. This would open new opportunities for high-precision
electroweak studies \cite{erler}. 

In order to study $e^{+}\,+\,e^{-}\,\rightarrow\,{\rm H}\,+\,Z$, 
$e^{+}\,+\,e^{-}\,\rightarrow\,t\bar{t}$ and $\gamma\,+\,\gamma\,\rightarrow\,
{\rm H}\,+\,{\rm H}$, we propose to install 70-MeV/m rf cavities in the same
tunnel once the technology for their production becomes available. The maximum
centre-of-mass energy would then be $2{\rm E}_{e} \approx 350$ GeV (${\rm E}_
{\gamma\gamma} \approx 280$ GeV), sufficiently high to produce $t\bar{t}$ 
pairs. From a scan of the $t\bar{t}$ production cross-section in the $t$-pair
threshold region, the top-quark mass could be mesured with $10^{2}$ MeV
accuracy.

\vspace*{1cm}
{\bf Erratum\,:}\hspace*{3mm}In Ref. \cite{belusev} on page 8, the sentence
beginning with ``The rich set of final states in $\gamma\gamma$, $e\gamma$ and
$e^{-}e^{-}$ collisions\,$\ldots$" should read ``The rich set of final states 
in $\gamma\gamma$, $pp$ and $e^{+}e^{-}$ collisions\,$\ldots$".

\end{document}